\documentclass[preprint,prd,amsmath,amssymb,amsthm,nofootinbib]{revtex4}
\usepackage[mathscr]{euscript}
\usepackage{bm}
\usepackage{graphicx}
\usepackage{subfigure}
\usepackage{appendix}
\usepackage{multirow}  
\usepackage{floatrow}
\usepackage{array}
\usepackage[colorlinks=true,linkcolor=red,urlcolor=blue]{hyperref}
\newcommand{\bea}{\begin{eqnarray}}
\newcommand{\eea}{\end{eqnarray}}
\newcommand{\beq}{\begin{equation}}
\newcommand{\eeq}{\end{equation}}

\def\/{\over}

\begin{document}
\title{Scalar induced gravitational waves in inflation with gravitationally enhanced friction}
\author{ Chengjie Fu,  Puxun Wu\footnote{Corresponding author: pxwu@hunnu.edu.cn} and Hongwei Yu  }
\affiliation{Department of Physics and Synergetic Innovation Center for Quantum Effects and Applications, Hunan Normal University, Changsha, Hunan 410081, China}

\begin{abstract}
We study the scalar induced  gravitational wave (GW) background in inflation with  gravitationally enhanced friction (GEF). The GEF mechanism, which is realized  by assuming a nonminimal derivative coupling between the inflaton field and gravity, is used to amplify the small-scale curvature perturbations to generate  a sizable amount of  primordial black holes. We find that the GW energy spectra   can reach  the detectable scopes of the future GW projects, and the power spectrum of curvature perturbations has a power-law form in the vicinity of  the peak. The scaling of  the GW spectrum  in the ultraviolet regions is two times that of the  power spectrum slope, and has a lower bound. In the infrared regions, the slope of the GW spectrum can be described roughly by a  log-dependent form. These features of the  GW spectrum may be used to check the GEF mechanism   if the scalar induced GWs are detected in the future.

\end{abstract}

\maketitle

\section{Introduction}
\label{sec1}
Primordial black holes (PBHs) may be formed in the very early era of the Universe.  Their implications for astronomy and cosmology have recently been receiving  extensive attention. It has been pointed out that PBHs can be taken as the possible sources of some astronomical events, on one hand. For example, the stellar-mass ($\sim\mathcal{O}(10)M_{\odot}$) PBHs are considered to be the promising candidate responsible for some gravitational wave (GW) events~\cite{Bird2016,Clesse2017,Sasaki2016}, which are radiated by  the binary black hole (BH) mergers and have been detected by the LIGO-Virgo Collaboration~\cite{LIGO}.  PBHs with earth mass ($\sim\mathcal{O}(10^{-5})M_{\odot}$)  can account for six ultrashort-timescale microlensing events in the 5-year OGLE data set~\cite{OGLE1, OGLE2}.  
In addition, an earth-mass PBH,  if it is captured by the Solar System,  has  been used to explain the anomalous orbits of trans-Neptunian objects~\cite{Planet9}. On the other hand, PBHs have also  been proposed as the candidate of dark matter, and they have a possibility for making up all dark matter in two asteroid-mass intervals:   $\mathcal{O}(10^{-16}) \sim (10^{-14})M_{\odot}$ and $\mathcal{O}(10^{-13}) \sim (10^{-11})M_{\odot}$~\cite{Katz2018,HSC2019, WD, EG,INTEGRAL1,INTEGRAL2}. 

The PBH formation  requires that the primordial curvature perturbations produced during the inflation era, which seed  the large-scale structure of the Universe, have a suitably large amplitude. Since the cosmic microwave background (CMB) observations have constrained the curvature perturbations to have an amplitude about $10^{-5}$ \cite{Planck} at the CMB (large) scale, in order to generate a sizable  amount of PBHs we need to enhance the small-scale curvature perturbations during inflation. When these enhanced curvature perturbations reenter the horizon, they result in the formation of black holes in the overdensed regions, and, at the same time,  lead to very large  scalar metric perturbations. These scalar   perturbations couple with the tensor perturbations at the second order, although they  decouple with each other at the linear level.  Inevitably, the large scalar metric perturbations become a significant GW source and generate abundant GW signals via the second-order effect to form the stochastic GW background~\cite{Kohri2018,RG2019_1,RG2019_2,Bartolo:2018evs,Bartolo:2018rku,Inomata2019,Wang2019,YF2019,Liu2019_1,Lu2019, Hajkarim}.\footnote{ It is worthing pointing out that the scalar induced GWs are gauge independent~\cite{Luca9689, Inomata00785,  Yuan00885}. } Therefore,  detecting the concomitant induced GW signals  provides an inspiring possibility and a whole new way to search for the existence of PBHs.

The mechanisms of amplifying the small-scale curvature perturbations within the inflationary scenario have been extensively studied in recent years \cite{YF2018,Chen2019,Ballesteros2019,Ozsoy18,Kamenshchik2019,Pi2018,Inomata2018,Tada2019,Liu2019_2,Bellido2017,Germani2017,Hu2017,Ezquiaga2018,Gong2018,
Ballesteros2018,Dalianis2019,Drees2019}. The most common    
 mechanism is the single-field inflection-point inflation \cite{Bellido2017,Germani2017,Hu2017,Ezquiaga2018,Gong2018, Ballesteros2018,Dalianis2019,Drees2019}, which requires that  the potential of the inflaton field has an approximate inflection point.  The rolling of the inflaton decelerates  when it comes close to the inflection point. As a result, the Universe experiences a period of ultraslow-roll inflation in the vicinity of  the inflection point, which enhances the curvature perturbations.

Recently, a novel enhancement mechanism of curvature perturbations was proposed by us in \cite{Fu2019}, in which the velocity of inflaton is decreased   by the increased  friction. To enhance the friction, we consider a nonminimal derivative coupling between the inflation field and gravity with the action being 
\begin{align}\label{action}
\mathcal{S}=\int d^4x \sqrt{-g}\left[\frac{1}{2\kappa^2}R - \frac{1}{2}\left(g^{\mu\nu}-\kappa^2 \theta(\phi) G^{\mu\nu}\right)\nabla_\mu\phi\nabla_\nu\phi - V(\phi)\right]\;,
\end{align}
where $\kappa^{-1}\equiv M_{\mathrm{pl}}=2.4\times10^{18}\;\mathrm{GeV}$ is the reduced Planck mass, $g$  the determinant of the metric tensor $g_{\mu\nu}$, $R$  the Ricci scalar, $G^{\mu\nu}$  the Einstein tensor, $\theta$  a dimensionless coupling parameter, which is function of $\phi$, and $V(\phi)$  the potential of the inflaton field.    By choosing  a special function form of   $ \theta(\phi)$, i.e., $\theta ={\omega}/{\sqrt{\kappa^2\left(\frac{\phi-\phi_c}{\sigma}\right)^2+1}}$,  a high-friction region can be realized at the slow-roll stage through the mechanism of the gravitationally enhanced friction (GEF) 
\cite{Fu2019,Germani2011_1,Germani2011_2,Tsujikawa2012}. Here, $\omega$, $\sigma$ and $\phi_c$ are constants. 
Thus, the inflaton evolves even more slowly in this high-friction region than in other slow-roll regions, which implies that a period of ultraslow-roll inflation is achieved when the inflaton goes through the high-friction region. The amplitude of the curvature perturbations is amplified during this ultraslow-roll inflation era.   In Ref. \cite{Fu2019}, we demonstrated that PBHs with physically attractive masses, such as $\mathcal{O}(10)M_{\odot}$, $\mathcal{O}(10^{-5})M_{\odot}$, $\mathcal{O}(10^{-12})M_{\odot}$, can be generated as a result of the amplified curvature perturbations by the GEF mechanism.

The PBH formation is accompanied inevitably  with the production of significant GW backgrounds induced by the overly large curvature perturbations. The detection of such GW signals may serve as  evidence  of  the GEF mechanism. This motivates us to carry out the investigation  in the present paper,  that is to perform a comprehensive analysis on the scalar induced GWs in the inflation model  with a nonminimal derivative coupling.
We organize our paper as follows: In Sec. II, we outline the basic formulas about the second-order GWs and calculate the GW energy spectra associated with the PHB formation studied in Ref. \cite{Fu2019}. In Sec. III, we analyze the scaling of the scalar induced GW spectra. Section IV gives our conclusions. In addition, the main equations of the nonminimal derivative coupling inflation model are outlined in the appendix.

\section{Gravitational waves induced by curvature perturbations}
\label{sec2}
We first give the basic formulas for  scalar induced GWs  during the radiation-dominated era through the second-order effect of the curvature perturbations. Ignoring the anisotropic stress, the perturbed Friedmann-Robertson-Walker (FRW) metric in the conformal Newtonian gauge has the form~\cite{Ananda2007}
\begin{align}
ds^2=a(\eta)^2\left\{ -(1+2\Psi)d\eta^2 +\left[(1-2\Psi)\delta_{ij}+\frac{h_{ij}}{2} \right]dx^idx^j \right\}\;,
\end{align}
where $a$ is the cosmic scale factor, $\eta\equiv \int a^{-1}dt$ is the conformal time, $\Psi$ is the first-order scalar perturbation, and $h_{ij}$ represents  the second-order transverse-traceless tensor perturbation.

After inflation, the inflaton will have decayed into light particles to thermalize our Universe once the reheating finishes. As a result,  when the Universe is dominated by the radiation, the inflation field has almost negligible effects on the cosmic evolution and can be neglected.   So, to investigate the scalar induced GWs during the radiation-dominated era, we only need to consider the standard Einstein equation 
and so the equation of motion for the second-order $h_{ij}$ satisfies
\begin{align}\label{EOM_GW}
h_{ij}^{\prime\prime}+2\mathcal{H}h_{ij}^\prime - \nabla^2 h_{ij}=-4\mathcal{T}^{lm}_{ij}S_{lm}\;,
\end{align}
where a prime denotes the derivative with respect to the conformal time and $\mathcal{H}\equiv a^{\prime}/a$ is the conformal Hubble parameter. 
 Here $\mathcal{T}^{lm}_{ij}$ is the transverse-traceless projection operator and $ S_{ij}$ is the GW source term~\cite{Ananda2007,Baumann2007}
\begin{align}
S_{ij}=4\Psi\partial_i\partial_j\Psi+2\partial_i\Psi\partial_j\Psi-\frac{1}{\mathcal{H}^2}\partial_i(\mathcal{H}\Psi+\Psi^\prime)\partial_j(\mathcal{H}\Psi+\Psi^\prime)\; .
\end{align}
At the radiation-dominated era, the scalar metric perturbation $\Psi$ in the Fourier space satisfies  the equation  
\begin{align}
\Psi_k^{\prime\prime}+ \frac{4}{\eta}\Psi_k^\prime+\frac{k^2}{3}\Psi_k=0\;, 
\end{align}
which admits  a solution~\cite{Baumann2007}
\begin{align}
\Psi_k(\eta)=\psi_k\frac{9}{(k\eta)^2}\left(\frac{\sin(k\eta/\sqrt{3})}{k\eta/\sqrt{3}}-\cos(k\eta/\sqrt{3}) \right)\;,
\end{align}
where $k$ is the comoving wave number, and $\psi_k$ is the primordial perturbation, which relates to the power spectrum of the curvature perturbation through  
\begin{align}
\langle \psi_{\bf k}\psi_{ \tilde{\bf k}}  \rangle = \frac{2\pi^2}{k^3}\left(\frac{4}{9}\mathcal{P}_\mathcal{R}(k)\right)\delta(\bf{k}+ \tilde{\bf k})\;.
\end{align}

During the radiation-dominated era, the energy density of the scalar induced GWs per logarithmic interval of $k$ can be evaluated as \cite{Kohri2018}
\begin{align}\label{OGW}
\Omega_{\rm{GW}}&(\eta_c,k) = \frac{1}{12} \int^\infty_0 dv \int^{|1+v|}_{|1-v|}du \left( \frac{4v^2-(1+v^2-u^2)^2}{4uv}\right)^2\mathcal{P}_\mathcal{R}(ku)\mathcal{P}_\mathcal{R}(kv)\nonumber\\
&\left( \frac{3}{4u^3v^3}\right)^2 (u^2+v^2-3)^2\nonumber\\ 
&\left\{\left[-4uv+(u^2+v^2-3) \ln\left| \frac{3-(u+v)^2}{3-(u-v)^2}\right| \right]^2  + \pi^2(u^2+v^2-3)^2\Theta(v+u-\sqrt{3})\right\}\;
\end{align}
at time $\eta_c$, which represents the time when $\Omega_{\rm{GW}}$ stops to grow. Here $\Theta$ is the Heaviside theta function. Between the energy spectra of the induced GWs at present and at $\eta_c$ there exists a relation~\cite{Inomata2019}
\begin{align}\label{OGW0}
\Omega_{\rm{GW},0}h^2 = 0.83\left( \frac{g_c}{10.75} \right)^{-1/3}\Omega_{\rm{r},0}h^2\Omega_{\rm{GW}}(\eta_c,k)\;,
\end{align}
where $\Omega_{\rm{r},0}h^2\simeq 4.2\times 10^{-5}$ is the current density parameter of radiation. Here $g_c\simeq106.75$ denotes the effective degrees of freedom in the energy density at $\eta_c$. 
 The  current frequency $f$ of the scalar induced GWs relates to the comoving wave number $k$  through the following equation 
\begin{align}\label{f}
f=1.546\times10^{-15}\frac{k}{1\rm{Mpc}^{-1}}\rm{Hz}\;.
\end{align}

\begin{figure}
\centering
\includegraphics[width=0.6\textwidth ]{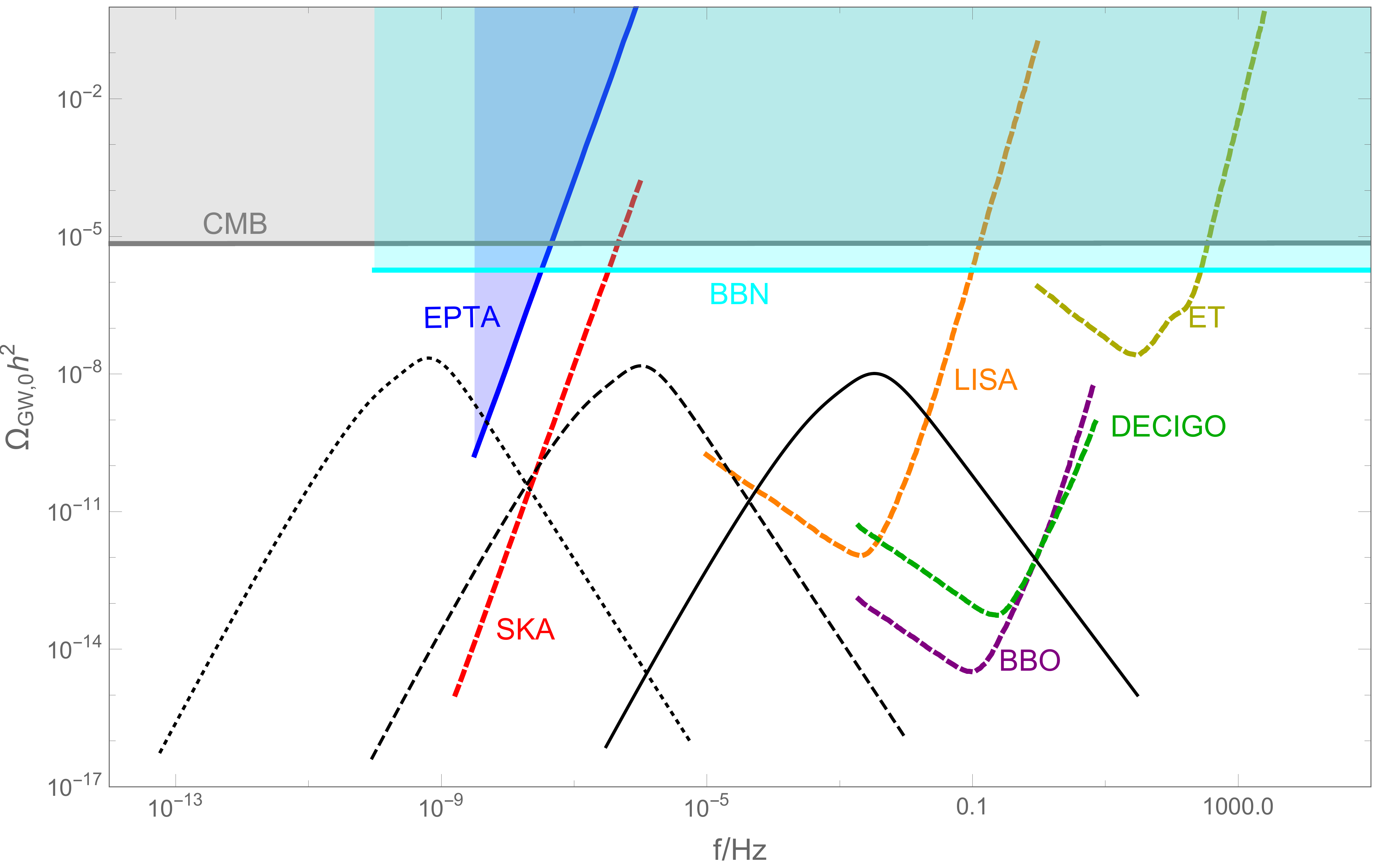}
\caption{\label{fig1} The current energy spectra of the induced GWs predicted by our model (black line). The solid/dashed/dotted black lines correspond to the production of asteroid-mass, earth-mass, and stellar-mass PBHs, respectively. The shaded regions represent the existing constraints on GWs from CMB \cite{Smith2006}, big bang nucleosynthesis \cite{Kohri2018}, and EPTA \cite{Lentati2015}. The other dashed lines are the expected sensitivity curve of the future gravitational-wave projects summarized in \cite{Moore12015}.} 
\end{figure}

Integrating Eq.~(\ref{OGW}) numerically  and using Eqs.~(\ref{OGW0}) and (\ref{f}), one can obtain the predicted current energy spectra of the scalar induced GWs associated with the production of PBHs. The results are shown in Fig.~\ref{fig1}, in which the solid, dashed and dotted curves correspond to the PBHs with asteroid mass, earth mass, and stellar mass, respectively. Apparently, the GW spectra have almost the same shape and amplitude, but  different peak frequencies. The smaller  the mass of PBHs, the higher the peak frequency of the GW spectrum. In the case of asteroid-mass PBHs, the peak of the GW spectrum is located in the sensitive region of LISA and the high-frequency part of the GW spectrum exceeds the sensitivity curves of deci-hertz interferometer GW observatory and big bang observer. The low- and high-frequency parts of the  GW spectrum related to the earth-mass PBH formation are above sensitivities of SKA and LISA respectively.  However, the dotted curve indicates that the scalar induced  GW spectrum associated with the stellar-mass BH generation crosses the regions excluded by european pulsar timing array (EPTA),  although the curvature perturbations generating stellar-mass PBHs  meet the EPTA constraint~\cite{Fu2019}. This constraint  is obtained  conservatively by parametrizing the power spectrum of curvature perturbations~\cite{Inomata2019}. In~\cite{Fu2019},  the Press-Schechter approach with the Gaussian window function has been used to discuss the production rate of PBHs.  It has been pointed in Ref. \cite{Tada2019} that if one adopts  the Press-Schechter approach with  the real-space top-hat window function instead of the Gaussian window function or the refined peak-theory approach~\cite{Germani2019} to calculate the abundance of PBHs, the required curvature perturbations are relatively smaller and then the corresponding GWs are consistent with the current EPTA constraint. Thus, we  guess that  the scalar induced GW from the GEF mechanism can also be consistent with the EPTA  if   the real-space top-hat window function  or the refined peak-theory approach is adopted. Besides the scalar induced GW spectra,  another important quantity  is the scaling of the  GWs since it is crucial to distinguishing different mechanisms for the generation of large primordial curvature perturbations.

\section{Scaling of scalar induced gravitational waves}
\label{sec3}
Since the shape of  scalar induced GW spectra relates to the scaling of the power spectrum of curvature perturbations, we  first derive the slope of the power spectrum  in the vicinity of the peak analytically and compare the analytical result with  that from  numerically solving  the Mukhanov-Sasaki equation.   After introducing a dimensionless time variable $\tau\equiv\sqrt{\lambda}\kappa^{-1}t$, where $\lambda$ is a  dimensionless parameter  in the inflationary potential~(\ref{A6}),  and a dimensionless field $\bar\phi\equiv\kappa\phi$, the Friedmann equation and the dynamical  equation of the inflaton field  given in Eqs.~(\ref{AEOM1}) and (\ref{AEOM3}) can be reexpressed as
\begin{align}
3\bar H^2 \simeq \bar\phi^{2/5}\;,
\end{align}
\begin{align}\label{EoM2}
\bigg(1+3 \bar\theta(\bar\phi)\bar H^2\bigg)\frac{d^2\bar\phi}{d\tau^2} + \bigg(1+3\bar\theta(\bar\phi)\bar H^2\bigg)3\bar H \frac{d\bar\phi}{d\tau}+\frac{3}{2}\bar\theta_{,\bar\phi}\bar H^2 \left(\frac{d\bar\phi}{d\tau}\right)^2 + \frac{2}{5}\bar\phi^{-3/5}\simeq 0\;,
\end{align}
where $\bar H \equiv (da/d\tau)a^{-1}$, $\bar\theta(\bar\phi)=\lambda \theta(\bar\phi)$ and $\bar\theta_{,\bar{\phi}}=d\bar\theta/{d\bar{\phi}}$ have the following forms
\begin{align}
\bar\theta = \frac{\sigma\omega\lambda}{\sqrt{(\bar\phi-\bar\phi_c)^2+\sigma^2}}\;,\qquad \bar\theta_{,\bar{\phi}}=-\frac{\sigma\omega\lambda(\bar\phi-\bar\phi_c)}{[(\bar\phi-\bar\phi_c)^2+\sigma^2]^{3/2}}\;.
\end{align}

\begin{figure}
\centering
\subfigure{\label{fig2a}}{\includegraphics[width=0.48\textwidth ]{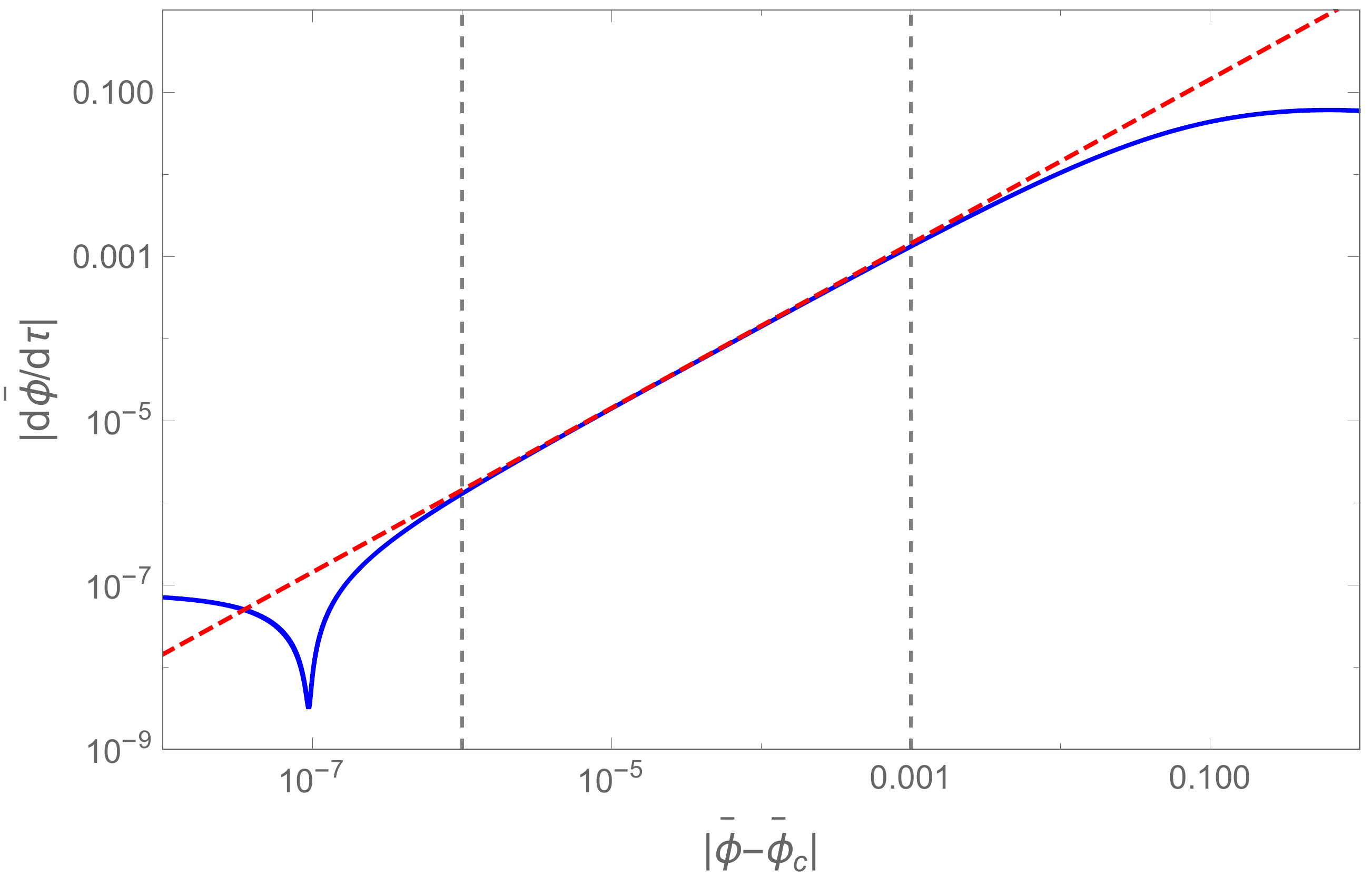}}
\subfigure{\label{fig2b}}{\includegraphics[width=0.48\textwidth ]{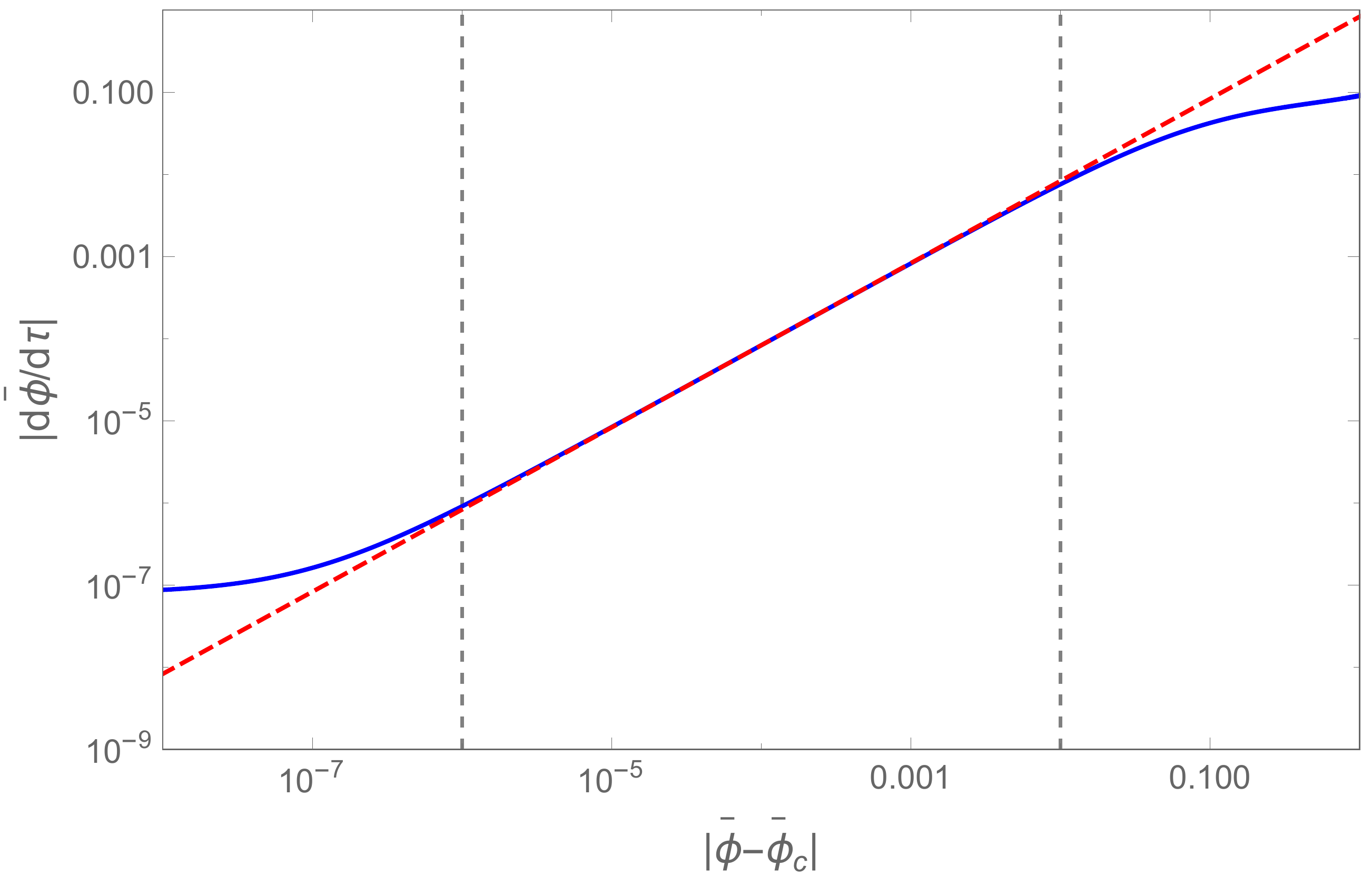}}
\caption{\label{fig2} The evolution of $|d\bar\phi/d\tau|$ as a function of $|\bar\phi-\bar\phi_c|$ when $\bar\phi>\bar\phi_c$ (left panel) and $\bar\phi<\bar\phi_c$ (right panel).  The parameters are set for the earth-mass PBH generation (the second row of Table~\ref{table1} in the appendix).  The blue solid lines represent the results by solving equations numerically and the red dashed lines are our analytical results given in Eq. (\ref{LS}). }
\end{figure}

Since the value of $\sigma$ ($\sim 10^{-9}$) is very very small, the value of $|\bar\phi-\bar\phi_c|$ is usually much larger than that of $\sigma$ except for the regions where $\bar\phi$ is extremely  close to $\bar\phi_c$.  So, we can investigate the ultraslow-roll inflationary dynamics under the condition $\sigma \ll |\bar\phi - \bar\phi_c|$.  In this case  $\bar\theta$ and $\bar\theta_{,\bar{\phi}}$ can be simplified to be
\begin{align}\label{theta}
\bar\theta\simeq\left\{\begin{array}{lcl}
+\frac{\sigma\omega\lambda}{\bar\phi-\bar\phi_c}\;,\;(\bar\phi>\bar\phi_c)\\
-\frac{\sigma\omega\lambda}{\bar\phi-\bar\phi_c}\;,\;(\bar\phi<\bar\phi_c)\end{array}\right.\;, \qquad
\bar\theta_{,\bar\phi}\simeq\left\{\begin{array}{lcl}
-\frac{\sigma\omega\lambda}{(\bar\phi-\bar\phi_c)^2}\;,\;(\bar\phi>\bar\phi_c)\\
+\frac{\sigma\omega\lambda}{(\bar\phi-\bar\phi_c)^2}\;,\;(\bar\phi<\bar\phi_c)\end{array}\right.\;.
\end{align}
 
In Eq.~(\ref{EoM2}), the second term in the left-hand side is the friction term arising from the derivative coupling and the cosmic expansion.  As discussed in Ref. \cite{Fu2019}, the strong friction condition $\bar\theta(\bar\phi)\bar H^2 \gg 1$ can be  satisfied easily in the vicinity of $\phi_c$.  In addition, $\bar H \simeq \bar\phi_c^{1/5}/\sqrt{3}$ is almost a constant    in the vicinity of $\phi_c$. Thus, during  the ultraslow-roll era  Eq.~(\ref{EoM2})  reduces to 
\begin{align}
\frac{\sigma\omega\lambda}{\bar\phi-\bar\phi_c}\frac{d^2\bar\phi}{d\tau^2}& + \sqrt{3}\bar\phi_c^{\frac{1}{5}}\frac{\sigma\omega\lambda}{\bar\phi-\bar\phi_c}\frac{d\bar\phi}{d\tau}-\frac{1}{2}\frac{\sigma\omega\lambda}{(\bar\phi-\bar\phi_c)^2}\left(\frac{d\bar\phi}{d\tau}\right)^2+\frac{2}{5}\bar\phi_c^{-1}\simeq 0 \;,\; \rm{for}\;\bar\phi>\bar\phi_c\;, \nonumber \\
-\frac{\sigma\omega\lambda}{\bar\phi-\bar\phi_c}\frac{d^2\bar\phi}{d\tau^2}& - \sqrt{3}\bar\phi_c^{\frac{1}{5}}\frac{\sigma\omega\lambda}{\bar\phi-\bar\phi_c}\frac{d\bar\phi}{d\tau}+\frac{1}{2}\frac{\sigma\omega\lambda}{(\bar\phi-\bar\phi_c)^2}\left(\frac{d\bar\phi}{d\tau}\right)^2+\frac{2}{5}\bar\phi_c^{-1}\simeq 0 \;,\; \rm{for}\;\bar\phi<\bar\phi_c\;.
\end{align}
This equation  has a linear solution:
\begin{align}\label{LS}
\frac{d\bar\phi}{d\tau}\simeq\left\{\begin{array}{lcl}
\sqrt{3}\bar\phi_c^{1/5}\left(\sqrt{1-\frac{4}{15}\bar\phi_c^{-7/5}(\sigma\omega\lambda)^{-1}}-1\right)(\bar\phi-\bar\phi_c)\;,\; (\bar\phi>\bar\phi_c)\\ \\
\sqrt{3}\bar\phi_c^{1/5}\left(\sqrt{1+\frac{4}{15}\bar\phi_c^{-7/5}(\sigma\omega\lambda)^{-1}}-1\right)(\bar\phi-\bar\phi_c)\;,\; (\bar\phi<\bar\phi_c)\end{array}\right.\;. 
\end{align}
Figure \ref{fig2} shows the evolution of $|d\bar\phi/d\tau|$ as a function of $|\bar\phi-\bar\phi_c|$ with the parameters set for the earth-mass PBH generation (the second row of Table~\ref{table1} in the appendix), which  also serves  as a concrete example in the subsequent numerical calculation  used  to test our analytic results.
It should be noted that the results in the other two cases of Table~\ref{table1} are similar to the  case of earth-mass PBH.
From this figure, one can see that the relations between $d\bar\phi/d\tau$ and $(\bar\phi-\bar\phi_c)$ given in Eq. (\ref{LS}) are nicely consistent with the numerical results.

\begin{figure}
\centering
\includegraphics[width=0.6\textwidth ]{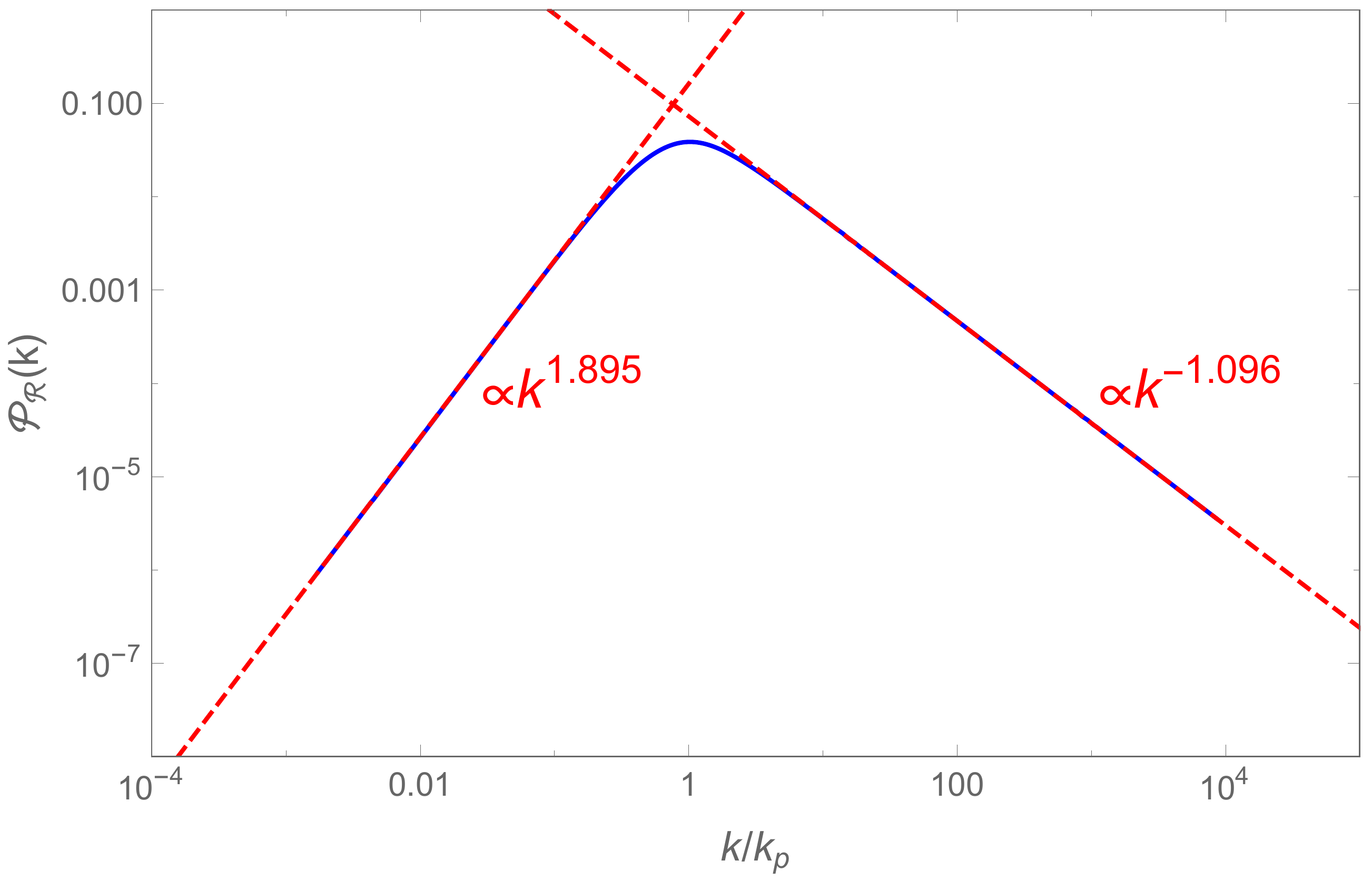}
\caption{\label{fig3} The power spectrum of curvature perturbations versus $k/k_p$ with the parameters  chosen to be the second row of Table~\ref{table1} in the appendix. The blue solid line is the power spectrum obtained by solving numerically the Mukhanov-Sasaki equation. Two red dashed lines have the slopes of $k^{1.895}$ and $k^{-1.096}$ as the analytic prediction given in Eq. (\ref{PL}) .} 
\end{figure}

\begin{figure}
\centering
\includegraphics[width=0.6\textwidth ]{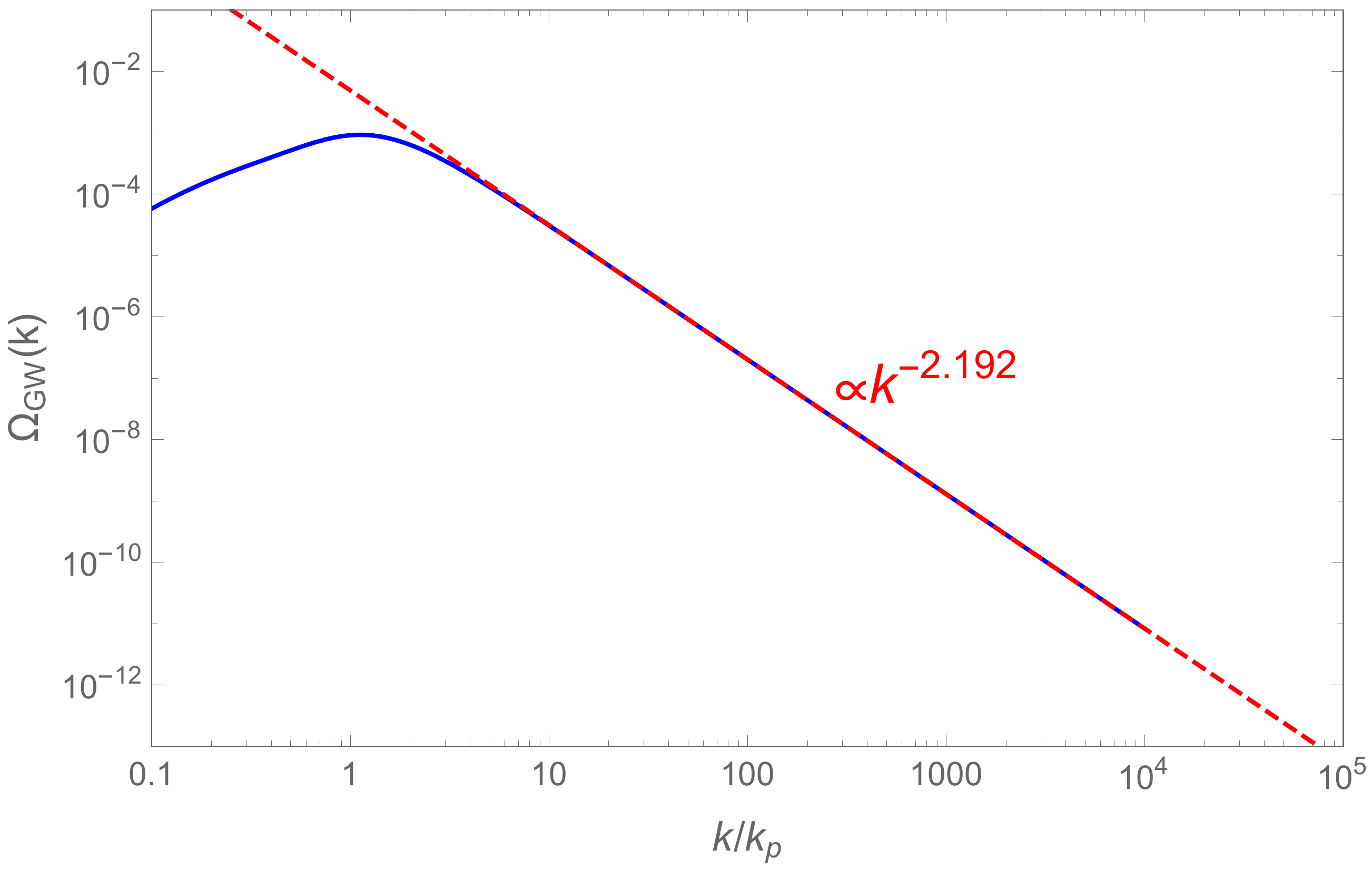}
\caption{\label{fig4} The energy spectrum of the induced GWs as a function of $k/k_p$ at $\eta_c$. The parameters are set to be the same as those in Fig.~\ref{fig3}. The solid blue line is the numerical result, and the dashed red line corresponds to the analytic result $\Omega_{\rm{GW}}\propto k^{-2.192}$.} 
\end{figure}

During the ultraslow-roll era that results from the GEF  ($\bar\theta(\bar\phi)\bar H^2 \gg 1$), the slow-roll conditions $\delta_X$ and $\delta_D$ defined in (\ref{SRC}) satisfy $\delta_X\ll \delta_D$. So,   Eq. (\ref{APR}) indicates  that  the power spectrum of curvature perturbations $\mathcal{P_R}$ is inversely proportional to $\delta_D$. Using Eqs.~(\ref{theta}) and (\ref{LS}),  one can obtain $\mathcal{P_R} \propto (\bar\phi-\bar\phi_c)^{-1}$. Since the tilt of the power spectrum is calculated to be
\begin{align}
\frac{d\ln\mathcal{P_R}}{d\ln k}=\left(\frac{d\ln\mathcal{P_R}}{dt}\right)\left(\frac{dt}{d\ln(aH)}\right) \simeq -\frac{1}{\bar H(\bar\phi-\bar\phi_c)}\frac{d\bar\phi}{d\tau}\;,
\end{align}
we obtain a power-law power spectrum for the curvature perturbations 
\begin{align}\label{PL}
\mathcal{P_R}\simeq\left\{\begin{array}{lcl}
k^{n_1}\;,\;(k<k_p)\\
k^{n_2}\;,\;(k>k_p)\end{array}\right.
\end{align}
by using Eq. (\ref{LS}). Here, the spectral indices  take the forms
\begin{align}\label{slope}
n_1=3\left(1-\sqrt{1-\frac{4}{15}(\kappa\phi_c)^{-7/5}(\sigma\omega\lambda)^{-1}}\right)\;,\nonumber \\
n_2=3\left(1-\sqrt{1+\frac{4}{15}(\kappa\phi_c)^{-7/5}(\sigma\omega\lambda)^{-1}}\right)\;.
\end{align}
and $k_p$ represents the comoving wave number corresponding to the peak of the power spectrum.  
The power spectra given in Eq.~(\ref{PL}) and  that obtained by numerically solving the Mukhanov-Sasaki equation (\ref{MS}) are compared in Fig.~\ref{fig3}, from which one can see that the power spectrum in our model can be well modeled by a power law with the slopes being $n_1$ and $n_2$ in the vicinity of peak. We must point out  that the expression of $n_1$ given in Eq.~(\ref{slope}) is valid only when $4(\kappa\phi_c)^{-7/5}(\sigma\omega\lambda)^{-1}\lesssim 15$. Actually, numerical calculations indicate that if $4(\kappa\phi_c)^{-7/5}(\sigma\omega\lambda)^{-1}> 15$, our mechanism of enhancing the curvature perturbations will become inefficient and the amplitude of the power spectrum cannot grow up to reach  the typical value expected to generate a sufficient abundance of PBHs. As a result, we give a rough bound on the slope of power spectrum with $n_1 \lesssim 3$ and $n_2 \gtrsim -1.24$.

\begin{figure}
\centering
\includegraphics[width=0.6\textwidth ]{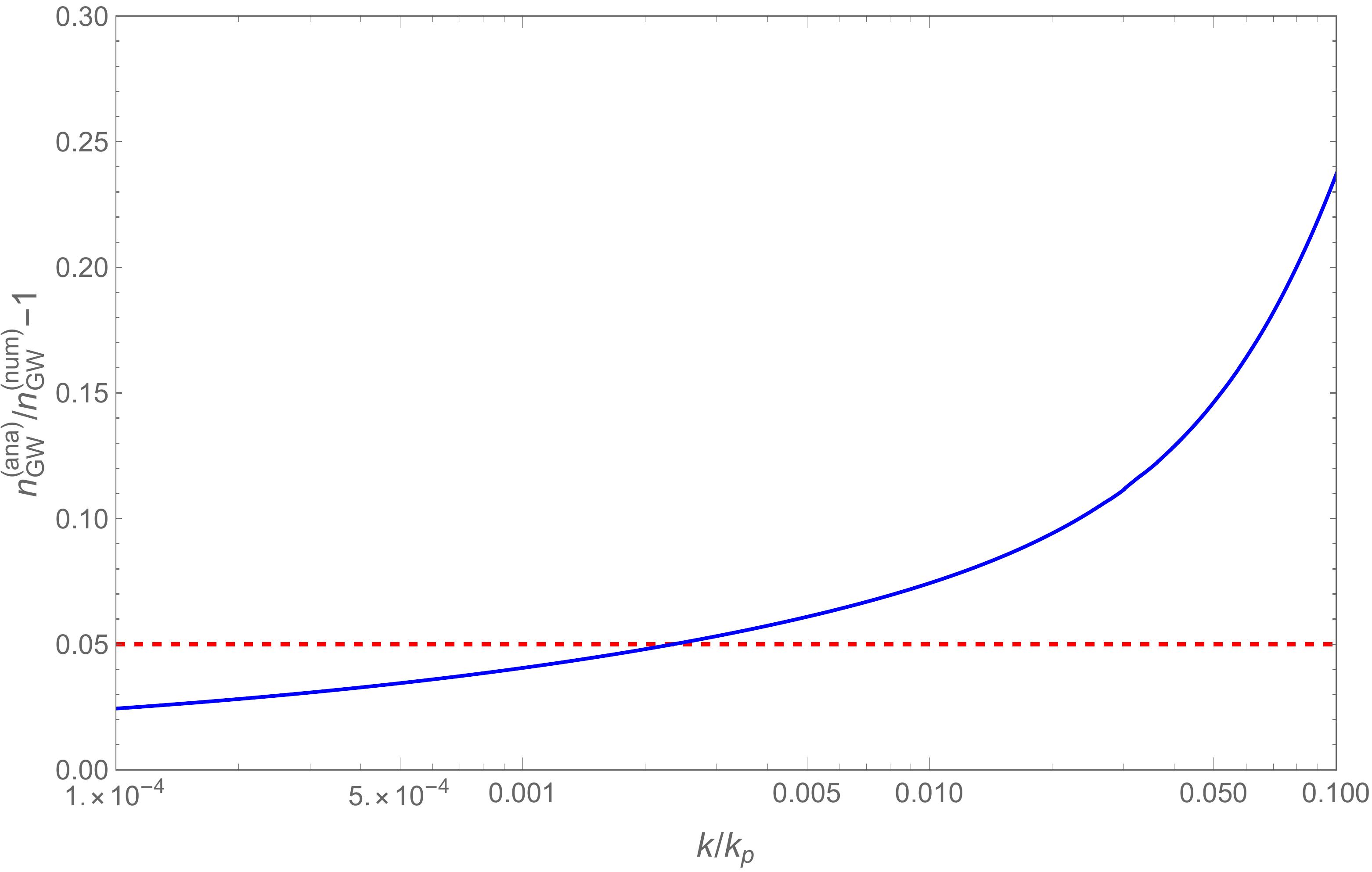}
\caption{\label{fig5} The relative error of $n_{\rm{GW}}$ as a function of $k/k_p$. $n_{\rm{GW}}^{(\rm{ana})}$ is the analytic result given in Eq. (\ref{log}) and $n_{\rm{GW}}^{(\rm{num})}$ is the numerical one. The parameters are set to be the same as those in Fig.~(\ref{fig3}).} 
\end{figure}

A comparison of Fig. \ref{fig4}  and \ref{fig3}   reveals that  in the ultraviolet regions ($k\gg k_p$) the power-law behavior of the scalar induced GWs has a slope $n_{\rm{GW}}\simeq 2n_2$, which is consistent with the result given in Ref.~\cite{Liu2019_1}  where it was found that if the power spectrum has the form  $\mathcal{P_R}\propto k^n$ with $n>-4$   the induced GW spectra have a $k^{2n}$ slope when  $k \gg k_p$.   Because of $n_2 \gtrsim -1.24$,  the slope of the induced GWs is roughly limited to be $n_{\rm{GW}} \gtrsim -2.48$ in the ultraviolet regions. However,  in the infrared regions ($k \ll k_p$), we find that the induced GW spectrum approaches to being nearly  scale invariant when $k / k_p$ is very very small, i.e. $k / k_p \lesssim 5\times 10^{-7}$, which is not plotted   in this paper since it is much less than the detectable scope of future GW experiments. For the regions $5\times 10^{-6} \lesssim k/k_p \ll 1$, we find that the scaling of the scalar induced GWs can be described  roughly by  an approximate log-dependent slope  given in~\cite{Yuan2019}  
\begin{align}\label{log}
n_{\rm{GW}} = 3 - \frac{4}{\ln{\frac{4k_p^2}{3k^2}}}\;.
\end{align}
In Fig. \ref{fig5}, we plot the relative error of $n_{\rm{GW}}$ for the model considered in the present paper by comparing the approximate and numerical results. With the decrease of $k/k_p$ the error becomes smaller and smaller,  which is less than $5\%$ when $k/k_p \lesssim 2\times10^{-3}$,  and  the GW spectrum approaches closer and closer  to the $k^3$ form obtained in \cite{CaiPi}.  

\section{conclusions}
\label{sec4}
We recently proposed a GEF mechanism to amplify the amplitude of curvature perturbations at  small scales and then found that a sizable amount of PBHs can be generated~\cite{Fu2019}. 
Following this work, we further discuss the production of the GWs induced by the enhanced curvature perturbations in this paper. We consider three typical GW spectra associated with the formation of PBHs with the mass around $\mathcal{O}(10)M_{\odot}$, $\mathcal{O}(10^{-5})M_{\odot}$ and $\mathcal{O}(10^{-12})M_{\odot}$, respectively. We find that the   GW signals induced by the curvature perturbations   can be probed by the future GW experiments.  However,   the energy spectrum of GWs, which relates to the stellar-mass PBH production estimated by using the Press-Schechter approach with the Gaussian window function, fails to satisfy the current constraint from EPTA. Our result  is in agreement with what was obtained in~\cite{Tada2019}, and moreover, the authors there also  point out that  once  the Press-Schechter approach with  the real-space top-hat window function or the refined peak-theory approach to calculate the abundance of PBHs is adopted, the required curvature perturbations are relatively smaller and then the corresponding GWs will be consistent with the current EPTA constraint. Thus, we guess that the incompatibility between the predicted GWs from the GEF mechanism  and the EPTA observation can also be avoided by using the real-space top-hat window function or the refined peak-theory approach for the PBH production.  Furthermore, we examine  the scaling  of the power spectrum of curvature perturbations and the scalar induced GW spectrum.  We find that, in the vicinity of  peak, the power spectrum has a power-law form. In the ultraviolet regions,  the scaling  of  the GW spectrum is two times  that of the  power spectrum slope, and has a lower bound. Whereas,   in the infrared regions, the slope of the GW spectrum can only be described roughly by a  log-dependent form. These features of the  GW spectrum from the amplified curvature perturbations may be used to check the GEF mechanism   if the scalar induced GWs are successfully detected in the future.

\begin{acknowledgments}

We thank Dr. Jing Liu and Dr. Shi Pi very much for fruitful discussions. This work was supported by the National Natural Science Foundation of China under Grants No. 11775077, No. 11435006, and No. 11690034, and by the Science and Technology Innovation Plan of Hunan province under Grant No. 2017XK2019.

\end{acknowledgments}

\appendix
\section{Main formulas of the inflation model with a nonminimal derivative coupling}
\label{appendix}
From the action (\ref{action}), we derive the following equations in the spatially flat FRW background,
\begin{align}\label{EOM1}
3H^2=\kappa^{2}\left[\frac{1}{2}\bigg(1+9\kappa^2 \theta(\phi)H^2\bigg)\dot\phi^2+V(\phi)\right]\;,
\end{align}
\begin{align}\label{EOM2}
-2\dot H=\kappa^2\left[\bigg(1+3\kappa^2\theta(\phi)H^2-\kappa^2\theta(\phi)\dot H\bigg)\dot\phi^2-\kappa^2\theta_{,\phi}H\dot \phi^3- 2\kappa^2\theta(\phi)H\dot\phi\ddot\phi\right]\;,
\end{align}
\begin{align}\label{EOM3}
\bigg(1+3\kappa^2 \theta(\phi)H^2\bigg)\ddot\phi + \bigg[1+\kappa^2\theta(\phi)\bigg(2\dot H+3H^2\bigg)\bigg]3H\dot\phi+\frac{3}{2}\kappa^2\theta_{,\phi}H^2\dot\phi^2+V_{,\phi}=0\;,
\end{align}
where $\theta_{,\phi}=d\theta/d\phi$, $V_{,\phi}=dV/d\phi$, $H\equiv\dot{a}/a$ is the Hubble parameter and a dot denotes the derivative with respect to the cosmic time. 
The slow-roll inflation is characterized by 
\begin{align} \label{SRC}
\epsilon\equiv-\frac{\dot H}{H^2}\ll 1\;,\quad  \delta_\phi\equiv \left|\frac{\ddot \phi}{H\dot\phi}\right|\ll 1\;,\quad \delta_X\equiv\frac{\kappa^2\dot\phi^2}{2H^2} \ll 1\;, \quad
\delta_D\equiv\frac{\kappa^4\theta\dot\phi^2}{4} \ll 1\;.
\end{align}
In Ref. \cite{Fu2019}, we consider a special functional form of $\theta(\phi)$,
\begin{align}
\theta =\frac{\omega}{\sqrt{\kappa^2\left(\frac{\phi-\phi_c}{\sigma}\right)^2+1}}\;,
\end{align}
and a fractional power-law potential 
\begin{align}\label{A6}
V = \lambda M_{\rm{pl}}^{18/5}|\phi|^{2/5}\;,
\end{align}
where $\lambda$, $\omega$ and $\sigma$ are  dimensionless parameters, and $\phi_c$ has the dimension of mass. Table \ref{table1} gives the choices of these parameters for successfully generating the stellar-mass, earth-mass, and asteroid-mass PBHs.

\begin{table}
\caption{The three parameter sets considered in Ref. \cite{Fu2019}.}
\begin{tabular}{>{\centering}p{3cm}>{\centering}p{2cm}>{\centering}p{2cm}>{\centering}p{2cm}}
\hline
\hline 
$\#$ & $\phi_{c}/M_\mathrm{pl}$ & $\omega\lambda$ & $\sigma$  \tabularnewline
\hline 

$\sim\mathcal{O}(10)M_\odot$ & $4.63$ & $1.33\times10^7$ & $2.6\times10^{-9}$  \tabularnewline

$\sim\mathcal{O}(10^{-5})M_\odot$ & $3.9$ & $1.53\times10^7$ & $3\times10^{-9}$   \tabularnewline

$\sim\mathcal{O}(10^{-12})M_\odot$ & $3.3$ & $1.978\times10^7$ & $3.4\times10^{-9}$  \tabularnewline
\hline 
\end{tabular}
\label{table1}
\end{table}

The Mukhanov-Sasaki equation is given by
\begin{align}\label{MS}
u_k^{''} + \left(c_s^2k^2-\frac{z^{''}}{z}\right)u_k=0\;,
\end{align}
where $z\equiv\sqrt{2Q_s}a$ and $u\equiv z\mathcal{R}$. Note that $\mathcal{R}$ is the comoving curvature perturbation, $Q_s$ and $c_s^2$ have the following forms
\begin{align}\label{Qs}
Q_s=\frac{w_1(4w_1w_3+9w_2^2)}{3w_2^2}\;,
\end{align}
\begin{align}\label{Cs2}
c_s^2=\frac{3(2w_1^2w_2H-w_2^2w_4+4w_1\dot w_1w_2-2w_1^2\dot w_2)}{w_1(4w_1w_3+9w_2^2)}\;,
\end{align}
with
\begin{align}\label{w}
\begin{split}
&w_1=M_\mathrm{pl}^2(1-2\delta_D)\;,\\
&w_2=2H M_\mathrm{pl}^2(1-6\delta_D)\;,\\
&w_3=-3H^2M_\mathrm{pl}^2(3-\delta_X-36\delta_D)\;,\\
&w_4=M_\mathrm{pl}^2(1+2\delta_D)\;.
\end{split}
\end{align}
Assuming that $\{\epsilon,\dot{c_s}/(Hc_s),\dot{Q_s}/(HQ_s)\}\simeq \rm{const}$ and taking the limit $\{\epsilon,\dot{c_s}/(Hc_s),\dot{Q_s}/(HQ_s)\}\ll 1$, the power spectrum of curvature perturbations can be calculated approximately at the horizon crossing [$c_sk=aH$] as \cite{Kobayashi2011}
\begin{align}\label{PR}
\mathcal{P}_\mathcal{R}=\frac{k^3}{2\pi^2}\left|\frac{u_k}{z}\right|^2 = \frac{H^2}{8\pi^2Q_s c_s^3}\;.
\end{align}

As we have pointed out in Ref. \cite{Fu2019}, although  the slow-roll condition $\delta_\phi\ll 1$ is violated at the period of ultraslow-roll inflation,  other slow-roll conditions $\{\epsilon,\delta_X,\delta_D\}\ll 1$ are valid. Thus, during the inflationary phase,   Eqs. (\ref{EOM1})-(\ref{EOM3}) can always be  simplified, respectively, to be
\begin{align}\label{AEOM1}
3H^2 \simeq \kappa^2 V(\phi)\;,
\end{align}
\begin{align}\label{AEOM2}
-2\dot H\simeq\kappa^2\left[\bigg(1+3\kappa^2\theta(\phi)H^2\bigg)\dot\phi^2-\kappa^2\theta_{,\phi}H\dot \phi^3- 2\kappa^2\theta(\phi)H\dot\phi\ddot\phi\right]\;,
\end{align}
\begin{align}\label{AEOM3}
\bigg(1+3\kappa^2 \theta(\phi)H^2\bigg)\ddot\phi + \bigg(1+3\kappa^2\theta(\phi)H^2\bigg)3H\dot\phi+\frac{3}{2}\kappa^2\theta_{,\phi}H^2\dot\phi^2+V_{,\phi}\simeq0\;.
\end{align}
Using $\{\epsilon,\delta_X,\delta_D\}\ll 1$ and Eq. (\ref{AEOM2}), we have 
\begin{align}
Q_s \simeq M_{\rm{pl}}^2(\delta_X+6\delta_D)\;, \quad c_s^2\simeq 1\;,
\end{align}
and accordingly, the power spectrum given in Eq. (\ref{PR}) can be approximately written as
\begin{align}\label{APR}
\mathcal{P}_\mathcal{R}\simeq \frac{H^2}{8\pi^2M_{\rm{pl}}^2(\delta_X+6\delta_D)}\;.
\end{align}
Since $\delta_X>0$ and $\delta_D>0$, and $c_s^2$ is close to 1 in the inflationary phase, it is easy to see that for the model considered in the present paper,  $Q_s>0$ and $c_s^2>0$, which are required to avoid ghost and gradient instabilities \cite{Kobayashi2011,Ema2015}, are always satisfied.

\end{document}